# Lithium doping's effects on the microstructural, dielectric, energy storage, optical and electrical properties of BaTi0.89Sn0.11O3 ceramics

Salma AYADH[1*], Salma TOUILI[1,2], Youness HADOUCH[1,2,3], Salma ELMOULOUA[1], M'barek AMJOUD[1], Daoud MEZZANE[1,2], Lahcen ESSALEH[1], Kateryna PUSHKAROVA[4], Zdravko KUTNJAK[3], Igor A. LUK'YANCHUK[2,4], Mimoun EL MARSSI[2]

*1 Laboratory of Innovative Materials, Energy and Sustainable Development (IMED), Cadi- Ayyad University, Faculty of Sciences and Technology, BP 549, Marrakech, Morocco.*
*2 Laboratory of Physics of Condensed Matter (LPMC), University of Picardie Jules Verne, Scientific Pole, 33 rue Saint-Leu, 80039 Amiens Cedex 1, France.*
*3 Condensed Matter Physics Department, Jožef Stefan Institute, Jamova Cesta 39, 1000 Ljubljana, Slovenia.*
*4 Department of Building Materials, Kyiv National University of Construction and Architecture, Kyiv 03680, Ukraine.*

*Corresponding author:

E-mail: *s.ayadh1802@gmail.com*, *s.ayadh.ced@uca.ac.ma*

ORCID: *https://orcid.org/0009-0004-6093-1680*

**Abstract**

This research study the sol–gel synthesis of lithium-doped barium stannate titanate $BaTi_{0.89}Sn_{0.11}O_3$ ($BTS_{11}$) and investigates how varying composition with lithium affect its structural, morphological, ferroelectric, optical, and electrical properties. The phase of the sol–gel prepared samples with different compositions $BaLi_xTi_{0.89}Sn_{0.11}O_3$ ($BL_xTS_{11}$), where x = 0%, 2%, 4%, 6%, 8, and 10%, was examined using X-ray diffraction (XRD), which showed the development of a pure perovskite structure. Furthermore, structural characterization was performed with Raman spectroscopy to identify changes in structure and chemical environment resulting from Li incorporation. The doping caused changes in the morphology and grain size of $BL_xTS_{11}$ pellets, as examined through SEM micrographs. These micrographs showed clear alterations in grain size that were inversely related to the lattice strain of $BL_xTS_{11}$. The samples were subsequently examined for their ferroelectric properties. All samples exhibited their maximum dielectric constant at approximately 45-50 °C. Furthermore, relationships between the chemical, structural, and morphological characteristics of the $BL_xTS_{11}$ ceramics and their energy storage capabilities were identified. Based on the P-E hysteresis behavior, the $BL_6TS_{11}$ ceramic (x=6%) ferroelectric exhibited superior energy-storage capabilities, featuring a recoverable energy-storage density of 225 mJ/cm$^3$ and an energy-storage efficiency of 60%. An examination of the material's optical and electrical properties revealed that Li doping led to a reduction in the band gap values and a considerable increase in resistivity.

## 1. Introduction

Energy resources are essential for modern society, fueling industrial growth, technological progress, and daily activities. However, the rising energy demand, combined with environmental challenges, underscores the need for sustainable and efficient energy storage solutions, high-performance catalysts for wastewater treatment, and superior water-splitting technologies. In the quest to replace toxic lead-based perovskites used in electrical power devices, recent research has focused on developing environmentally friendly ferroelectric materials.

One of the most widely studied ferroelectric materials is Barium Titanate ($BaTiO_3$), a perovskite-type compound known for its non-toxic nature. Over the past decades, it has garnered significant attention due to its remarkable dielectric, ferroelectric, and piezoelectric properties, making it highly suitable for applications in capacitors, energy harvesting, electrocatalysis, sensors, and actuators [1], [2], [3]. However, researchers continue to explore ways to enhance its performance, particularly through doping strategies.

By incorporating isovalent or heterovalent dopants into either the A-site (Ba) or B-site (Ti) of $BaTiO_3$, key material properties can be significantly tuned. These modifications can improve temperature stability, Curie temperature, polarization response, electromechanical behavior, electrocatalytic efficiency, and energy storage capacity [4] [5][6]. Various doping systems have been explored, including $BaLi_xTiO_3$ [7], $Ba_{0.85}Ca_{0.15}Zr_{0.1}Ti_{0.9}O_3$ [8], $Ba(Ti_{0.99-1.25x}Mn_{0.01}Nb_x)O_3$ [9], $Ba_{0.9}Sr_{0.1}Ti_{1-x}Sn_xO_3$ [10], $(Ba_{1-x}Ca_x)(Ti_{0.85+x}Zr_{0.02}Sn_{0.13-x})O_3$ [11], $NaMg_{0.12}Nb_{0.88}O_{3-x}BiYbO_3$ [12], and $(1-x)(0.99\ Na_{0.5}Bi_{0.5}TiO_3-0.01BiYbO_3)-xSrTiO_3$ [13]. These studies have demonstrated enhanced physicochemical properties, particularly in energy storage, by increasing maximum polarization intensity and reducing residual polarization. Furthermore, optimizing doping levels has shown potential for further improving energy storage density by increasing breakdown strength, as reported in materials like $Sr_{0.6}(Bi_{0.5}Na_{0.5})_{0.4}TiO_3$, $Na_{0.5}Bi_{0.5}TiO_3-BiYbO_3-SrTiO_3$, and $Ba_{0.91}Ca_{0.09}TiO_3$ [14], [15], [16].

Among these modified perovskites, $BaTi_{0.89}Sn_{0.11}O_3$ ($BTS_{11}$) has demonstrated unique advantages. This ceramic feature double morphotropic phase boundaries that coexist near room temperature, resulting in a high dielectric constant, elevated piezoelectric coefficient, and improved energy storage, energy harvesting, and motion-sensing capabilities [17] [18]. Despite the promising effects of Sn doping in $BTS_{11}$, limited studies have explored the impact of other dopants on its properties [19].

To address this gap, the present study investigates for the first time the effects of lithium (Li) doping on the structural, dielectric, energy storage, and optical properties of $BTS_{11}$ ceramics.

## 2. Experimental

### 2.1 Synthesis Method

The sol–gel method was used to synthesize $BaLi_xTi_{0.89}Sn_{0.11}O_3$ with varying lithium concentrations (x = 0%, 2%, 4%, 6%, 8%, and 10%). The starting materials included barium acetate ($Ba(CH_3COO)_2$), tin chloride ($SnCl_2 \cdot 2H_2O$), titanium isopropoxide, and lithium acetate dihydrate ($LiC_2H_3O_2 \cdot 2H_2O$), serving as the precursors for barium, tin, titanium, and lithium, respectively.

Initially, barium acetate and lithium acetate were dissolved in acetic acid in the required stoichiometric proportions to form Solution 1. Meanwhile, tin chloride was separately dissolved in 2-methoxyethanol and stirred for 1 hour. Following this, a stoichiometric volume of titanium isopropoxide was introduced into the tin chloride solution, which was then stirred for another 1 hour, forming Solution 2.

Next, Solutions 1 and 2 were combined at room temperature and thoroughly stirred. To achieve a clear and transparent solution, ammonia was carefully added dropwise. The mixture was then heated to form a gel, which was subsequently dried at 100 °C and calcined at 1000 °C for 4 hours to obtain a fine powder. The resulting calcined powder was further ground for 1 hour, mixed with 5% polyvinyl alcohol (PVA) for granulation, and uniaxially pressed into ceramic pellets.

Finally, the $BL_xTS11$ pellets underwent sintering at 1250 °C for 7 hours in air, following the methodology established in previous studies [19]. The process flowchart is illustrated in Figure 1.

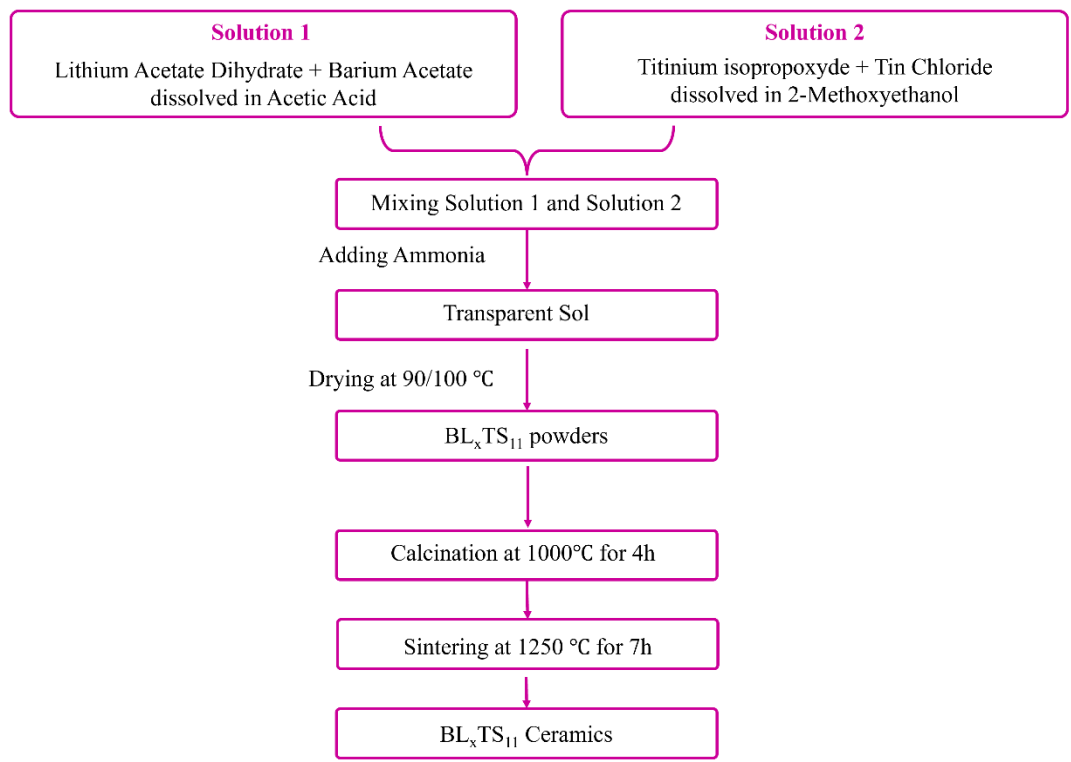

*Figure 1. The synthesis process of $BL_xTS_{11}$ ceramics by the sol-gel method.*

### 2.2. Instrumentation

The structural characterization of the ceramics was conducted using X-ray diffraction (XRD) with a Rigaku XRD diffractometer, employing Cu-Kα radiation (λ = 1.54 Å). Raman spectra were obtained using a Renishaw Raman spectrometer equipped with a CCD detector. The morphological analysis was performed through scanning electron microscopy (SEM) using an FEI Quanta 200 microscope.

The polarization-electric field (P–E) hysteresis loops were measured using a CPE1701 system (PloyK, USA), powered by a Trek 609-6 high-voltage supply (USA), with the samples immersed in a silicone oil bath at room temperature. The optical properties were analyzed using UV-Visible diffuse reflectance spectroscopy, recorded with a Shimadzu UV-2600 spectrophotometer.

## 3. Results and discussion

### 3.1. Microstructural characterization

As previously discussed, $BTS_{11}$ exhibits a near quadruple-phase coexistence at room temperature, where cubic, tetragonal, orthorhombic, and rhombohedral phases are present simultaneously. The XRD patterns of both undoped and Li-doped $BTS_{11}$ ceramics, shown in Figure 2a, confirm that all samples maintain a single-phase $BaTiO_3$ structure without detectable secondary phases. This suggests that $Li^+$ ions have successfully incorporated into the BT lattice.

A closer view of the XRD peaks around $2\theta = 45°$, depicted in Figure 2b, reveals a single prominent peak, indicating the dominant pseudo-cubic phase in $BTS_{11}$ at room temperature. The merging of the (200) and (002) reflections within the 44–46° range supports this structural characteristic. Additionally, the presence of asymmetric peaks suggests the potential coexistence of multiple phases. Wang et al. [20] proposed that, due to the comparable total energies of polar tetragonal, rhombohedral, and orthorhombic phases and the non-polar cubic phase, polar domains may randomly form and become embedded within the cubic phase matrix.

Figure 2c illustrates the relationship between Li doping concentration (x), cell volume, and XRD peak positions. As the Li content increases, the diffraction peaks shift, confirming the integration of Li into the $BTS_{11}$ crystal structure. For $x < 0.06$, lattice contraction occurs, shifting the (200) peak toward higher diffraction angles ($2\theta$), which reduces both the lattice parameter and the unit cell volume. However, beyond $x = 0.06$, an opposite trend emerges, with peaks shifting toward lower angles, leading to an expansion of the unit cell.

This variation in cell volume is strongly influenced by the specific crystallographic site occupied by $Li^+$ ions. When present in the A-site, Li incorporation causes a reduction in unit cell volume due to the smaller ionic radius of $Li^+$ (1.18 Å) compared to $Ba^{2+}$ (1.61 Å) in 12-fold coordination [22, 23]. Conversely, when $Li^+$ ions substitute at the B-site, the unit cell expands since $Li^+$ (0.76 Å, 6-fold coordination) is larger than $Ti^{4+}$ (0.61 Å) [24]. Initially, at lower doping levels ($x \leq 0.06$), $Li^+$ primarily occupies the A-site, while at higher concentrations, it progressively substitutes at the B-site, contributing to the observed increase in cell volume [21].

The calculated cell volume using UnitCell software and the corresponding shifts in XRD $2\theta$ positions with respect to Li doping levels are displayed in Figure 2c.

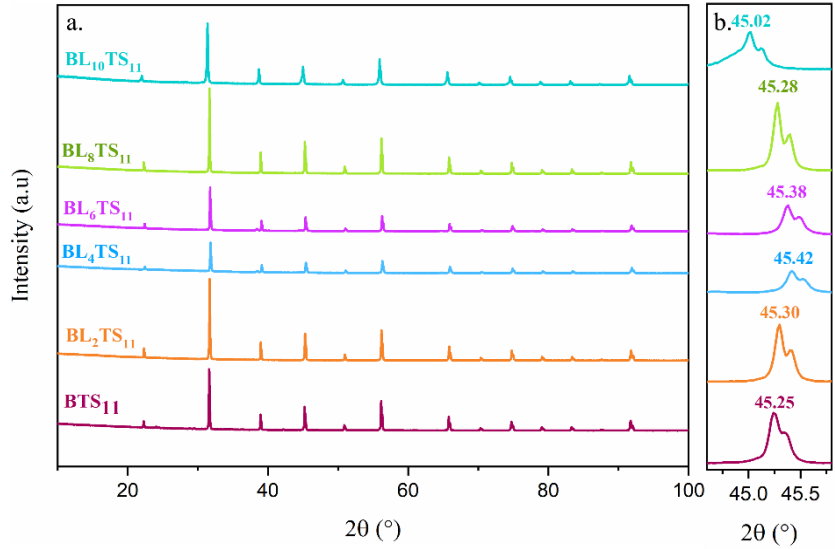

*Figure 2(a, b). Room temperature XRD pattern of $Ba_{1-x}Li_xTi_{0.89}Sn_{0.11}O_3$ ceramics.*

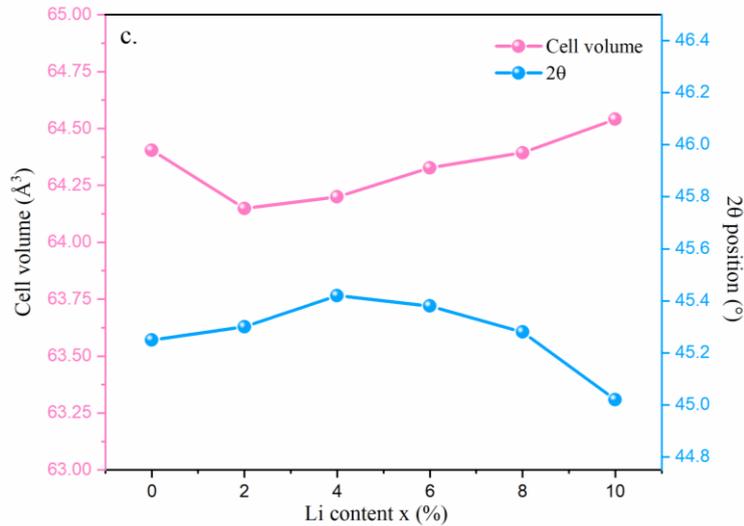

*Figure 2c. Variation of Cell volume and XRD 2θ position v.s Li ions content.*

Several properties, such as reduced polarization, the broadening and flattening of the temperature-dependent permittivity maximum, a lower dielectric constant and Curie temperature, as well as high tunability, have been attributed to stress-induced structural distortions associated with a significant reduction in grain size [25, 26]. Given this, the effect of Li concentration on both grain size and lattice strain was analyzed.

Figure 3 and 4 presents SEM micrographs and grain size distribution histograms for $BL_xTS_{11}$ ceramics, respectively. A clear trend is observed: the average grain size decreases from 1.15 μm (pure $BTS_{11}$) to 0.32 μm ($BL_6TS_{11}$) as $Li^+$ content increases, followed by an increase to 0.9 μm ($BL_8TS_{11}$) and 1.29 μm ($BL_{10}TS_{11}$). This suggests that the incorporation of $Li^+$ ions play a role in suppressing grain growth, likely due to their tendency to accumulate at grain boundaries, thereby limiting grain mobility [27]. A reduction in grain boundary mobility weakens mass transport, leading to smaller grains. The smallest grain size, observed for $BL_6TS_{11}$, is further linked to its high lattice strain [28]. Beyond this concentration, the increase in grain size is attributed to the formation of oxygen vacancies, which enhance mass transport during sintering, thereby facilitating grain growth [29].

Additionally, the lattice strain of $BL_xTS_{11}$, determined using the Scherrer method (Equation 1), exhibits a strong dependence on Li doping and follows a four-stage pattern (Figure 5):

$$\varepsilon = \beta/4\tan\theta \quad (eq.1)$$

where:

- $\varepsilon$ is the lattice strain,
- $\beta$ is the FWHM of the peak in radians,
- $\theta$ is the Bragg angle.

**Stage 1 (0%–2% Li doping):** Lattice strain decreases, likely because $Li^+$ ions, with their smaller ionic radius, integrate into the crystal lattice without causing major structural distortions.

**Stage 2 (2%–4% Li doping):** A gradual increase in lattice strain suggests a higher degree of Li substitution.

**Stage 3 (4%–6% Li doping):** A more pronounced increase in lattice strain is observed, likely due to the replacement of $Ti^{4+}$ (0.61 Å) by larger $Li^+$ ions (0.76 Å), which is known to induce strain [30].

**Stage 4 (8%–10% Li doping):** A reduction in lattice strain is noted, possibly due to the presence of charged oxygen vacancies, which may lead to lattice contraction [28].

A clear inverse relationship between grain size and lattice strain is observed, with $BL_6TS_{11}$ exhibiting the smallest grain size and highest strain. This behavior is likely associated with an increase in dislocation density, as higher lattice strain can generate more dislocations, ultimately restricting grain growth [28]. The initial trend ($BTS_{11}$ to $BL_2TS_{11}$) may be influenced by the type of dislocation present within the material. Specifically, dislocations moving across grains can be annihilated by oppositely signed dislocations at the grain boundary, reducing both dislocation density and lattice strain [31].

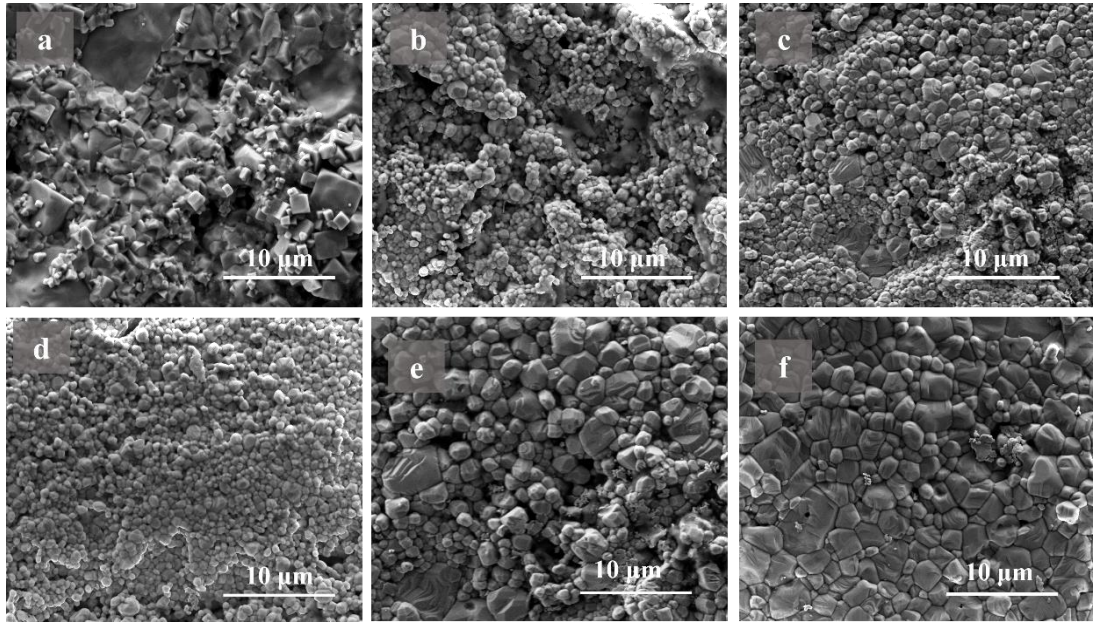

*Figure 3. SEM micrographs of the ceramics: **a.** $BTS_{11}$ **b.** $BL_2TS_{11}$ **c.** $BL_4TS_{11}$ **d.** $BL_6TS_{11}$ **e.** $BL_8TS_{11}$ **f.** $BL_{10}TS_{11}$.*

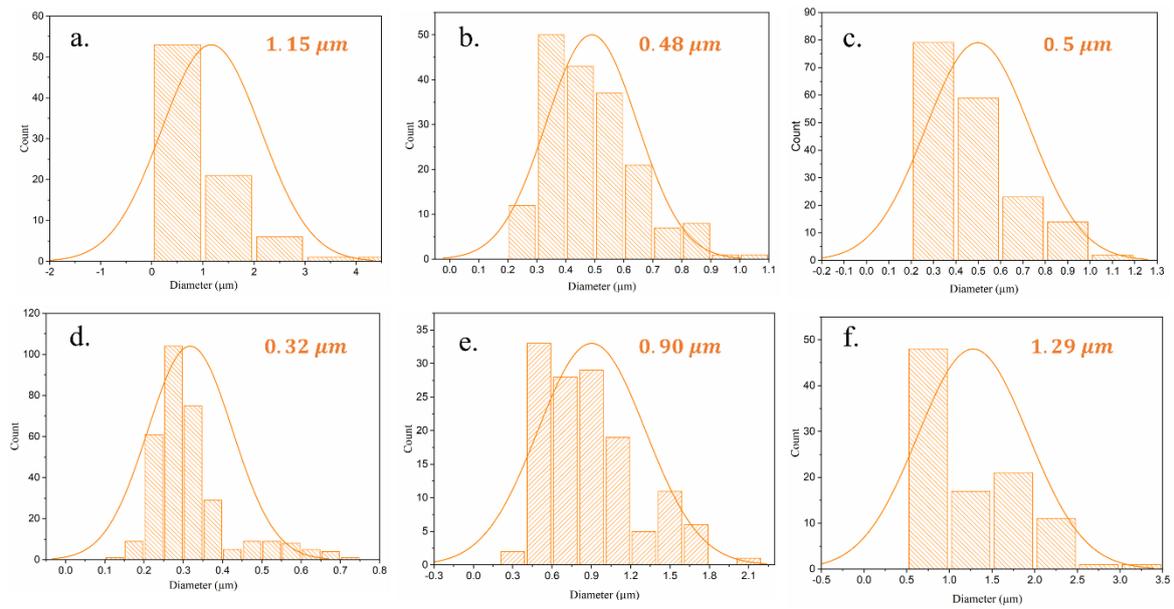

*Figure 4. Grain size distribution histograms of the ceramics: **a.** $BTS_{11}$ **b.** $BL_2TS_{11}$ **c.** $BL_4TS_{11}$ **d.** $BL_6TS_{11}$ **e.** $BL_8TS_{11}$ **f.** $BL_{10}TS_{11}$.*

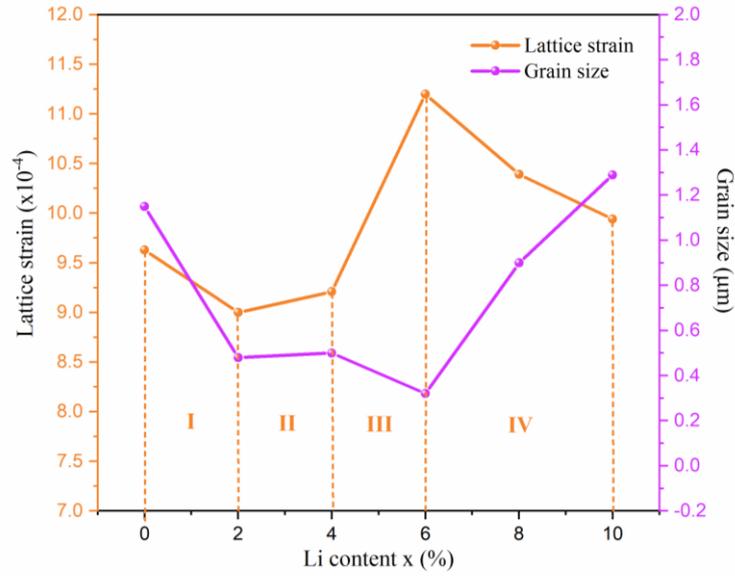

*Figure 5. Variation Lattice strain and grain size vs Li content.*

Since XRD analysis provides structural information based on a large volume of unit cells (exceeding 10 nm³), its precision in detecting mixed phases within a material is somewhat limited. This necessitates complementary techniques for a more detailed structural assessment. Raman spectroscopy, known for its sensitivity to short-range vibrational modes, serves as an effective tool for identifying polymorphic variations that may go undetected by XRD [32].

The Raman spectra of the synthesized $BL_xTS_{11}$ ceramics are illustrated in Figure 6a. In this context, any shifts in peak positions correspond to modifications in the chemical environment, localized stresses, and grain size effects [33]. The spectra reveal five distinct peaks at 188, 250, 305, 515, and 720 cm$^{-1}$, which are characteristic of the tetragonal $BaTiO_3$ phase [34]. The observed vibrational modes can be categorized into different spectral regions:

- Part I features modes associated with A-site vibrations and B-O chain interactions (frequencies below 200 cm$^{-1}$).
- Parts II and III contain modes attributed to the oxygen vibrations within the $BO_6$ octahedra in the $ABO_3$ perovskite framework [35].

Specifically, the peak at ~188 cm$^{-1}$ corresponds to the $A_1(TO)$ and $E(LO)$ modes, while the ~250 cm$^{-1}$ peak is assigned to the $A_1(TO)$ mode. A strong band appearing at approximately 515 cm$^{-1}$ is attributed to the transverse optical $A_1(TO)$ phonon mode, associated with $TiO_6$ octahedral vibrations in the tetragonal and/or cubic phase. Meanwhile, the weak peak at ~720 cm$^{-1}$ is linked to the highest-frequency longitudinal optical (LO) mode, characteristic of the tetragonal $BaTiO_3$ structure. These spectral features confirm the coexistence of cubic and tetragonal phases in $BL_xTS_{11}$ [36].

As depicted in Figure 6b, when Li content in the $BTS_{11}$ lattice remains below 6%, the Raman modes broaden, indicating an increase in structural disorder due to the incorporation of Li$^+$ ions, which have a different valence state and ionic radius compared to the host ions [37]. However, at 6% Li doping, the Raman peaks become sharper and more intense. This phenomenon is likely a result of high lattice strain associated with small grain size and the presence of mixed phases, both of which contribute to enhanced tetragonality. This is particularly evident in the narrow full-width at half-maximum (FWHM) values of the peaks at ~305 cm$^{-1}$ and ~720 cm$^{-1}$ for $BL_6TS_{11}$ [38].

Furthermore, the peak positions exhibit a systematic shift:

- For x < 6%, they shift toward higher frequencies, implying stronger chemical bonds.
- For x > 6%, they shift toward lower frequencies, suggesting a weakening of chemical bonds [39].

The downward frequency shift supports the hypothesis that Li$^+$ ions preferentially occupy the B-site. Since Li$^+$ (0.76 Å, six-fold coordination) has a larger ionic radius than Ti$^{4+}$ (0.61 Å), its incorporation may lead to distortions and shrinkage of the TiO$_6$ octahedra, consequently weakening bond strength within the crystal lattice [39]. Additionally, this shift has been associated with lattice expansion and the formation of oxygen vacancies, further contributing to structural modifications [40].

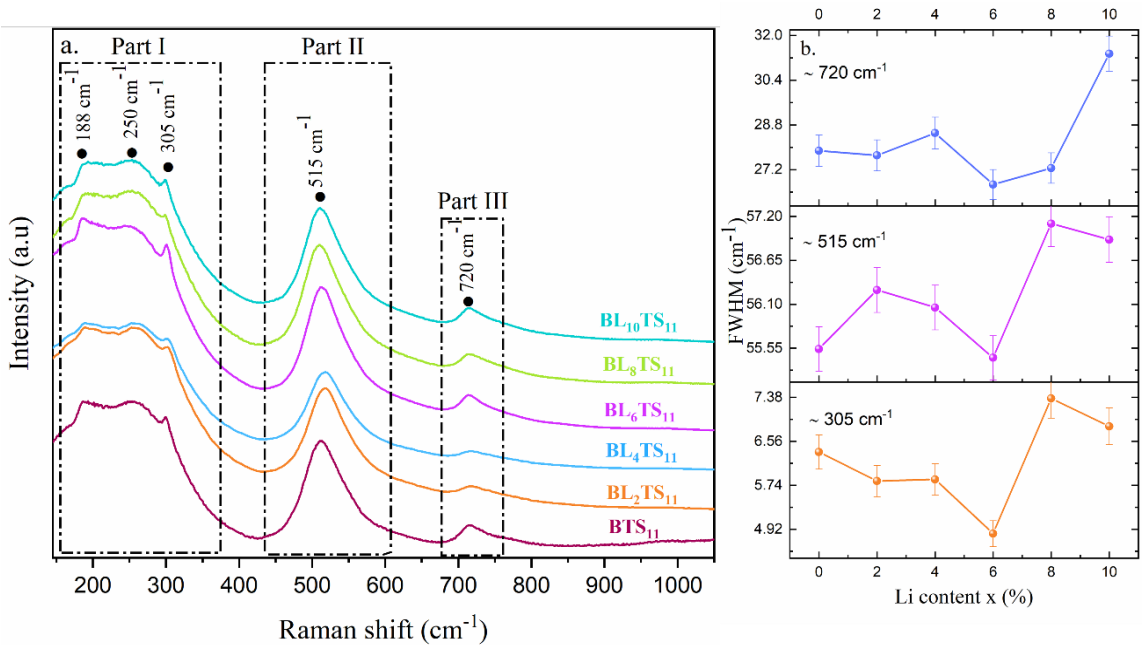

*Figure 6.* **a.** *Raman spectra and FWHM of $BL_xTS_{11}$ ceramics and variation of wavenumbers at 515 and 720 cm$^{-1}$ with Li concentration* **b**. *Variation of FWHM at wavenumber 305, 515 and 720 cm$^{-1}$ with Li content.*

### 3.3 Dielectric Properties

Figure 7a presents the temperature-dependent variation of the dielectric constant for $Ba_{1-x}Li_xTi_{0.89}Sn_{0.11}$ ceramics, measured at 100 kHz. The maximum permittivity ($T_{nax}$), which varies with Li content, is observed at $T_{nax} = 40°C$ for all samples. However, with an increase in Li concentration to 10%, a slight shift of $T_{nax}$ to higher temperatures is noted. This shift can be attributed to an enhanced interaction between Ti$^{4+}$ and O$^{2-}$ ions as a result of B-site substitution in BTS$_{11}$ [24].

Additionally, the presence of a single anomaly centered at ~40°C suggests the coexistence of multiple phases over a broad temperature range (20–65°C), aligning with the findings from XRD and Raman spectroscopy [17].

Previous studies indicate that the increase in dielectric permittivity observed for 2% and 4% Li-doped $BTS_{11}$ may be linked to their small grain size [41]. However, once the critical grain size is reached, further grain refinement contributes to a reduction in the dielectric constant [41]. This explains why $BL_6TS_{11}$ (6% Li doping) exhibits the lowest dielectric constant—the increase in grain boundary density leads to a decrease in relative permittivity ($\varepsilon_r$). This behavior is consistent with the composite model of dielectric ceramics, where grain cores typically have high permittivity, while grain boundaries exhibit low permittivity [42].

Furthermore, the broadening of the dielectric peak for $BL_6TS_{11}$ indicates a diffuse-type phase transition [43]. Beyond 6% Li doping, a second increase in the dielectric constant, particularly at 8% and 10% Li, can be attributed to the formation of oxygen vacancies induced by substitution effects, as suggested by Raman and SEM results. These oxygen vacancies generate charged defects within the crystal lattice, enhancing electronic polarization, which ultimately increases the dielectric constant [43].

Figure 7b illustrates the temperature dependence of dielectric loss ($\delta$) for different Li compositions. The $\delta$ values decrease gradually with increasing temperature and then stabilize at higher temperatures, indicating a thermally stable dielectric response across the measured range.

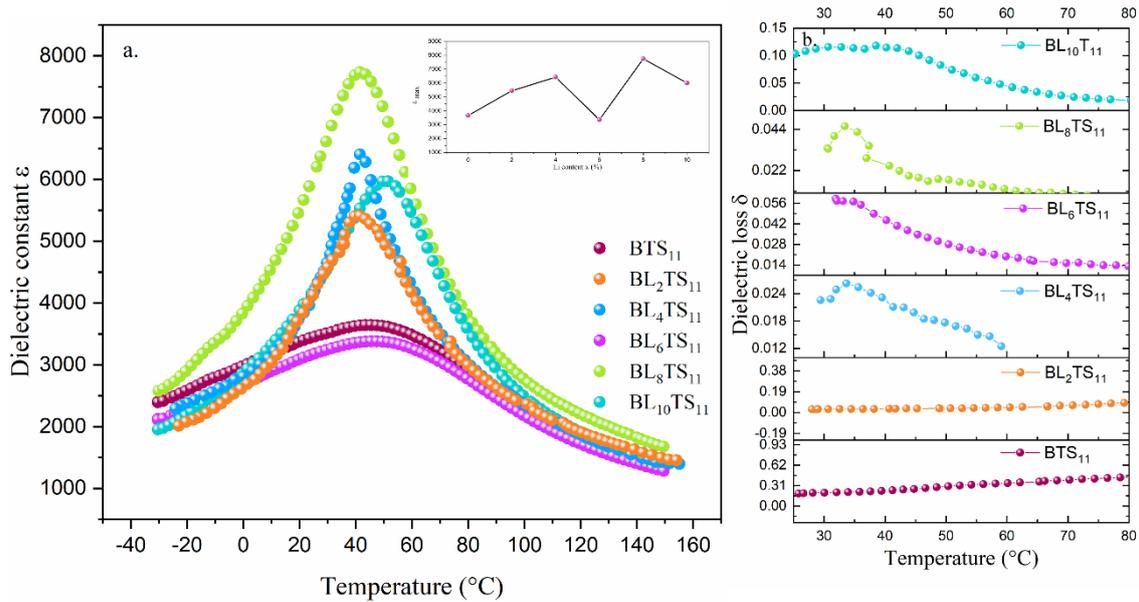

*Figure 7*. The temperature dependence of **a.** dielectric constant, and **b.** dielectric loss for $BL_xTS_{11}$ ceramics.

### 3.4 Ferroelectric properties and Energy storage performances

Figure 8a presents the polarization-electric field (P-E) hysteresis loops of Li-doped $BL_xTS_{11}$ ceramics. The loops exhibit the characteristic behavior of ferroelectric materials, becoming slimmer as the Li content (x) increases. The optimum energy-storage properties were observed in $BL_6TS_{11}$ (x = 6%), with a recoverable energy-storage density ($W_{rec}$) of 224.7 mJ/cm³ and an energy-storage efficiency ($\eta$) of 60.4%.

A key factor contributing to this enhanced energy density is the high lattice strain in $BL_6TS_{11}$, which may promote a greater degree of alignment of defect dipoles within the material (Figure 8b). This alignment is believed to enhance energy density, as suggested in previous studies [44, 45]. However, it is important to note that Xuetian Gong's study [46] presents a contrasting view, suggesting that reducing lattice strain improves energy storage. This discrepancy highlights the complex and non-trivial relationship between lattice strain and energy storage, emphasizing the need for further research to fully understand its effects.

Moreover, despite its relatively low dielectric constant, the energy-storage capability of a material is not solely determined by its permittivity. Other factors, such as microstructure and composition, also play a crucial role. In particular, a homogeneous microstructure with fine grain size, as evidenced by SEM results for $BL_6TS_{11}$, may contribute to an increase in breakdown strength (BDS), as reported in various studies [45, 47]. A high BDS allows the material to withstand stronger electric fields, thereby improving its overall energy density. Specifically, at 6% Li doping, a high electric field (E) of 64 kV/cm was achieved, further enhancing BDS and recoverable energy $W_{rec}$.

The values of remnant polarization ($P_r$) and coercive field ($E_C$), determined from the hysteresis loops, are summarized in Table 1. In general, $P_r$ decreases with increasing Li doping, except for $BL_2TS_{11}$, where a visible increase in polarization is observed. The decline in Pr can be attributed to the introduction of heterovalent $Li^+$ ions, which disrupt the long-range ferroelectric order in $BTS_{11}$ and promote the formation of polar nanoregions (PNRs) with higher dynamic activity, leading to a reduction in $P_r$ [48].

The increase in $P_r$ for $BL_2TS_{11}$ can be understood in terms of the type of vacancies formed when $Li^+$ is incorporated into the $BTS_{11}$ lattice. The charge imbalance introduced by $Li^+$ doping may be compensated by the formation of A-site ($Ba^{2+}$) vacancies, which reduce local constraint stress against polarization reversal, thereby increasing $P_r$ [49].

Furthermore, $BL_{10}TS_{11}$ exhibits the smallest coercive field ($E_C$), the lowest remanent polarization ($P_r$), and the highest energy-storage efficiency (η). This behavior can be linked to the presence of oxygen vacancies, as acceptor doping leads to the formation of defect dipoles involving doped ions and oxygen vacancies $[Li_{Ti} - V_O]^-$ [50]. The small $E_C$ value is likely due to the weakened interaction between defect dipoles and spontaneous polarization within the ferroelectric domains, which facilitates domain switching. Additionally, these defect dipoles can contribute to the reduction of remanent polarization ($P_r$) by allowing nanodomains to return to random orientations after the external electric field is removed [50].

*Table 1. Total energy ($W_{tot}$), Energy stored ($W_{rec}$), Energy efficiency (η), Coercive field ($E_C$) and Remanent polarization ($P_r$) values for all $BL_xTS_{11}$ ceramics.*

| BLxTS11 | $W_{tot}$ (mJ/cm³) | $W_{rec}$ (mJ/cm³) | η (%) | $E_C$ (kV/cm) | $E_{max}$ (kV/cm) | Pr (μC/cm²) | $P_s$ (μC/cm²) |
|---|---|---|---|---|---|---|---|
| *x=0.00* | 376.7 | **183.1** | 48.61 | 2.35 | 41.00 | 6.47 | 20.82 |
| *x=0.02* | 529.3 | **191.8** | 36.24 | 2.67 | 56.00 | 8.11 | 22.28 |
| *x=0.04* | 323.9 | **192.4** | 59.39 | 1.80 | 57.00 | 3.50 | 13.52 |
| *x=0.06* | 371.7 | **224.7** | 60.44 | 2.97 | 64.00 | 2.65 | 13.40 |
| *x=0.08* | 295.3 | **161.0** | 54.52 | 2.63 | 49.00 | 2.86 | 14.01 |
| *x=0.10* | 221.4 | **151.4** | 68.09 | 0.84 | 45.00 | 2.02 | 14.70 |

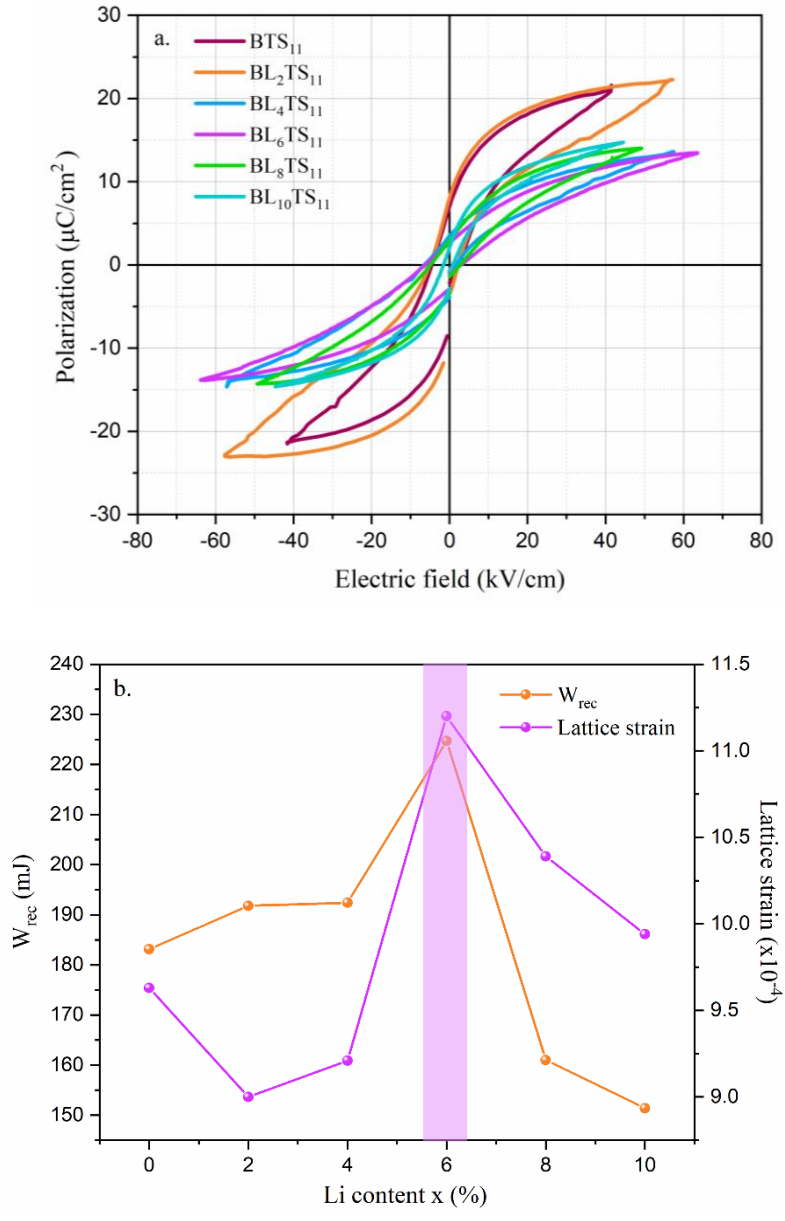

*Figure 8*. **a.** *P–E loops, and* **b.** *Correlation between $W_{Rec}$ and Lattice strain.*

**3.5 Optical and Electrical properties**

Due to their remarkable optical and piezoelectric properties, barium titanate ($BaTiO_3$) and its related materials have been widely used as lead-free piezoelectric materials for various applications, including water splitting, bacterial disinfection, and wastewater treatment [51]. To determine the optical band gap of the prepared materials, optical diffuse reflectance spectra were employed to investigate their optical characteristics. The band gap of the $BL_xTS11$ materials was determined using the Kubelka–Munk function and the Tauc equation (eq. 2):

$$F(R) = \frac{[(1-R)^2]}{2R} \text{ and } (\alpha h\nu)^{1/n} = A(h\nu - Eg) \quad (eq.2)$$

where R is the reflectance, hv is the incident photon energy, A is a proportionality constant, and n is an exponent dependent on the transition type. n = 1/2 corresponds to allowed direct transitions, while n = 2, 1.5, and 3 correspond to indirect transitions and forbidden transitions. $BaTiO_3$ and Sn-doped $BaTiO_3$ are direct band-gap materials, so the data were plotted using n = 2 [53].

By plotting the curve $(\alpha hv)^2$ vs hv, the band gap energy is determined by extrapolating the linear segment at high energies, and the band gap is identified on the hv axis in electron-volts (eV). The calculated band gap values for $BL_xTS_{11}$ are 3.34, 3.36, 3.48, 3.59, 2.09, and 2.50 eV for $BTS_{11}$, $BL_2TS_{11}$, $BL_4TS_{11}$, $BL_6TS_{11}$, $BL_8TS_{11}$ and $BL_{10}TS_{11}$ respectively.

It was observed that the band gap values generally increase with increasing Li content. However, for $BL_8TS_{11}$ and $BL_{10}TS_{11}$, the band gap decreases, as shown in Figure 9. This behavior can be attributed to the reduction in dielectric loss related to polarization at grain boundaries. As the grain size increases, the number of grain boundaries decreases, leading to a decrease in dielectric loss, which results in a wider band gap. The decrease in the band gap for BL8TS11 and BL10TS11 may further confirm the presence of oxygen vacancies, which create defect levels close to the conduction band, causing interactions that lower the band gap [54].

Additionally, the pure $BTS_{11}$ absorbs light at energies greater than 3.34 eV, indicating that it absorbs only UV light. However, the Li-doped ceramics ($BL_xTS_{11}$) with x ≥ 6% absorb in the visible light range, which could enhance their performance in catalytic degradation of pollutants in water and $H_2$ generation by combining the effects of photocatalysis under visible light and piezocatalysis in a single catalyst.

The electrical properties were also investigated by examining the resistivity of the $BL_xTS_{11}$ ceramics (Figure 9). The resistivity was examined using dielectric measurements, and the real and imaginary parts of the impedance (Z' and Z") were calculated from the dielectric parameters (ε' and ε") using the following formulas:

$$Z' = \frac{\varepsilon''}{((C_0 w)(\varepsilon'^2 + \varepsilon''^2))} \quad \text{and} \quad Z'' = \frac{-\varepsilon'}{((C_0 w)(\varepsilon'^2 + \varepsilon''^2))} \quad (eq.3)$$

where:

- ε' is the dielectric constant,
- ε" is the dielectric loss,
- $C_0$ is the vacuum capacitance,
- ω is the angular frequency.

Then, the resistivity (σ) was calculated using the following formula:

$$\sigma = \frac{e}{S} \cdot \frac{Z'}{Z'^2 + Z''^2} \quad (eq.4)$$

where:

- e is the thickness of the sample,
- S is the surface area of the sample.

The results show that Li+ doping significantly increased the resistivity, which may be due to the generation of lattice defects caused by the stoichiometric mismatch between $Li^{+}$, $Ba^{2+}$, $Ti^{4+}$, and $Sn^{4+}$ in $BTS_{11}$ [55].

As shown in Figure 8, the resistivity increases with the widening of the band gap. This is because, as the band gap becomes wider, the material becomes less conductive and more insulating.

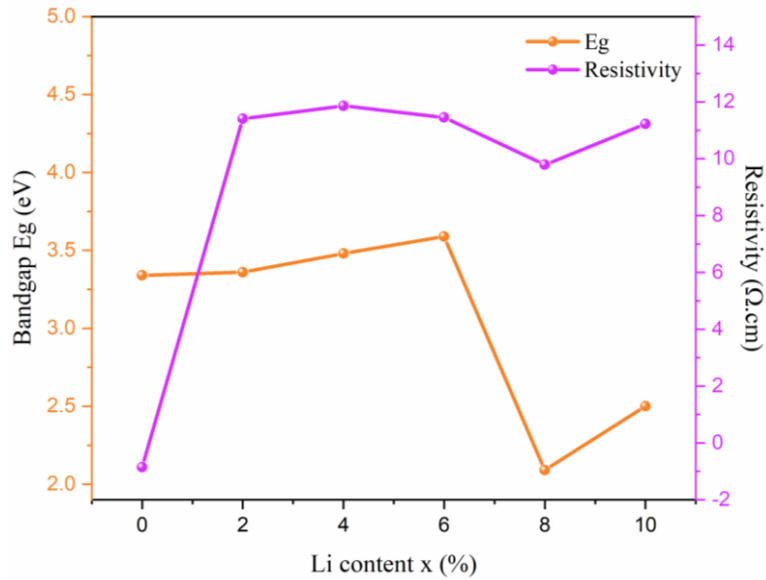

*Figure 9.* *Variation of band gap values determined from Tauc's plot of the transformed Kubelka-Munk function $[F(R)h\nu]^2$ against photon energy (hν) and Resistivity of $BL_xTS_{11}$ as a function of Li concentration.*

## 4. Conclusions

This study investigates, the effect of Li concentration (0%, 2%, 4%, 6%, 8%, and 10%) on $BaTi_{0.89}Sn_{0.11}O_3$ ($BTS_{11}$) ceramics synthesized using the sol-gel method. Various characterization techniques, including X-ray diffraction (XRD), Raman spectroscopy, and scanning electron microscopy (SEM), were employed to assess the structural, morphological, ferroelectric, and optical properties of the synthesized materials. The XRD and Raman analyses confirmed the successful formation of $BL_xTS_{11}$ with a single perovskite phase containing a mixture of cubic and tetragonal phases. These results indicated that the Li dopant entered the BTS11 crystal structure, inducing various modifications. It was suggested that Li first occupies the A-site, followed by the B-site of the $BTS_{11}$ structure. The Li doping was found to refine the crystal grain size, and at higher Li concentrations (8% and 10%), grain growth was promoted, which could be attributed to the formation of oxygen vacancies.

The sample with 6% Li concentration ($BL_6TS_{11}$) showed the smallest grain size due to high lattice strain. Dielectric measurements revealed that all samples exhibited the maximum permittivity within a temperature range of 45-50°C, with a single anomaly suggesting the coexistence of multiple phases. Additionally, the energy density was calculated for all samples, and it was found that the 6% Li-doped sample showed the highest energy density of 224.7 mJ/cm³ and an energy efficiency of 60%. This performance was attributed to the material's ability to withstand a high electric field (Emax = 64 kV/cm) compared to other compositions, due to an improvement in the breakdown strength (BDS).

Optical properties revealed an increase in band gap values and a corresponding rise in resistivity as the material became more insulating. However, the Li-doped $BL_xTS_{11}$ ceramics with x ≥ 6% exhibited a smaller band gap, suggesting the potential for broader applications. These materials could be useful for novel applications, including catalytic degradation of pollutants in water and hydrogen generation, in addition to their use in energy storage.

**Acknowledgements**

This work was supported by the H-GREEN project (No 101130520) and the European Union Horizon 2020 Research and Innovation Action MSCA-RISE-MELON (No. 872631). The Slovenian Research and Innovation Agency is acknowledged (research core funding P2-0105 and research project N2-0212).